\documentclass[a4paper,prb,showpacs,superscriptaddress]{revtex4}    

\usepackage{epsfig}

\begin{document}

\title{Hystersis like behaviour in Thin Films with heating-cooling cycle.}

\author{P. Arun}
 \affiliation{Department of Physics \& Astrophysics, University of
Delhi, Delhi - 110 007, India}
  \email{arunp92@physics.du.ac.in, arunp92@yahoo.co.in, agni@physics.du.ac.in}
\author{A.G. Vedeshwar}
 \affiliation{Department of Physics \& Astrophysics, University of
Delhi, Delhi - 110 007, India}

\begin{abstract}
The expression of temperature distribution along a film thickness is derived
and distribution of temperature in the film as the substrate is heated
is shown. The variation of film resistance with different substrate temperature
is calculated and the existence of temperature gradient along the film
thickness with finite thermal conductivity leads to hysteresis like
behaviour on heating-cooling the film. 
\end{abstract}

\pacs{73.61; 73.61.G; 81.40.C}

\maketitle

\section{Introduction}

The electrical conductivity measurements are important in characterising
conducting or semiconducting materials, both in their thin film and bulk 
state. It is routinely carried out for various materials. The temperature
dependence of resistivity yields information about intrinsic band gap of 
the material, the activation energy for conduction in films due to grain  
boundary barrier height or impurity activation energy etc. For the above  
estimation, the resistance measurement are taken either in the heating or 
cooling direction of temperature variation. If the system is heated or    
cooled very slowly, i.e. with same rate of change of temperature, both     
the heating and cooling cycles coincide. However, considerable difference
has been observed when an amorphous film is heated above crystalline 
transition temperature and cooled back to room temperature. This can       
be understood as due to structural changes\cite{1, 2, 3}. In such cases after    
cooling, most of the films do not regain their initial resistance. However 
some did regain their initial resistance, thus enclosing an area as in     
hysteresis loops. Hysteresis have been observed in bismuth films even      
without structural changes\cite{4} where the heating and cooling rates were    
kept different. This is interesting and since no attempt is made to explain
this variation, in this manuscript we attempt to explain the appearance of
hysteresis due to non-equilibrium state of the film.

\section{Theory}

We consider the film to be kept on a copper block which is heated by a 
heating coil embedded in it. The heating rate is varied by the voltage 
applied to the heating coil, such that the whole surface of the copper 
block is having uniform temperature. The film is kept on this copper block 
resulting in heating from the substrate side (fig 1). The temperature varies 
along the film thickness with time. The variation of temperature with time 
and spatial co-ordinates is given by\cite{5}
\begin{eqnarray}
c_v{\partial T \over \partial t} = \lambda {\partial^2 T \over \partial x^2}
\end{eqnarray}
where ${\lambda}$ is the thermal conductivity of the film and ${\rm c_v}$ is 
the specific heat of the film. A solution of this partial differential equation 
depends on the initial and boundary conditions of the problem. Depending on the 
initial and boundary conditions solution would be different\cite{6}. The initial 
condition for the film of thickness 'd' being heated from the substrate side 
would be given as 
\begin{eqnarray}
T(x=0, t=0) = T_{sub} \\
T(x=d, t=0) =  T_{sur} 
\end{eqnarray}
while after a long time the film would be uniformly heated, with the
surface attaining the same temperature as that of the substrate, i.e. 
\begin{eqnarray}
T(x, \infty) = T_{sub}\nonumber
\end{eqnarray}
From the given conditions, we search for a solution of the form
\begin{eqnarray}
T(x,t) =  g(x)h(t)\nonumber
\end{eqnarray}
Or in other words the variables are separable and thus, solving equation (1) by
separable variable method we have the solution 
\begin{eqnarray}
T(x,t)= a + b sin \left ( {\pi x \over 2 d} \right ) e^{-{\pi^2 D t \over 4
d^2}} 
\end{eqnarray}
where D is the thermal diffusivity, ${\rm \lambda / c_v }$. Applying the 
conditions stated in (2) and (3) the above solution may be written as
\begin{eqnarray}
T(x,t)= T_{sub} - (T_{sub}- T_{sur})sin\left ( \pi x \over 2 d \right ) 
e^{-{\pi^2 D t \over 4 d^2}} 
\end{eqnarray}
Under experimental conditions, where resistivity or resistance is measured as a
function of temperature, the substrate temperature would be continuously changing
with time i.e. would be time dependent. Hence the above equation can not be
used as it is. Numerically, first the surface temperature is calculated for a
given substrate temperature at a given instant, along with temperatures along 
the thickness of the film. The new surface temperature can be plugged back into
the expression along with the new substrate temperature. This scenario is valid
for film's with moderate thermal conductivity. Too carry out such a numerical
calculation we assume the substrate temperature to vary with time during the 
heating cycle as
\begin{eqnarray}
T_{sub}(t) = P(1-e^{-Qt})+R\nonumber
\end{eqnarray}
where 'R' is the room temperature. We have numerically determined the 
distribution of temperature along the thickness
of 1000\AA\, films of  varying diffusivity, as described in the last section.
Figure 2 shows the variation of temperature along the thickness of three 
different  diffusivities (a) ${\rm 5 \times 10^{-3} \AA^2/sec}$, (b) ${\rm 5
\times 10^2 \AA^2/sec}$ and ( c) ${\rm 5 \times 10^3 \AA^2/sec}$.
Diffusivity, as already stated is the ratio of thermal conductivity to the
specific heat of the material. Comparing three different diffusivity of same
thickness implies different materials of same thickness is being studied. If 
we assume there is not much variation in specific heat, the comparative study 
is being done for materials of varying thermal conductivity (l). For numerical 
computation we assume the values of the constants of  equation (1) to be
${\rm 360^oC}$, ${\rm 0.00039sec^{-1}}$ and ${\rm 14.5^oC}$ for P, Q
and R respectively. The saturation temperature that can be attained would be
${\rm \sim 380^oC}$, which would be very high. Hence, we assume the heater is 
switched off after 800sec, by which time, the copper block would be at ${\rm
\sim 110^oC}$. The family of curves show the spatial distribution of 
temperature at various given time, namely after (i) 0 sec, (ii) 200sec,
(iii) 400sec, (iv) 600sec and (v) 800 sec of heating. As is evident from
figure 2(a) the surface remains at room temperature since heat does not spread 
to the surface though the substrate is getting hotter with time. This is due to 
the poor thermal conductivity of the film. This difference in  surface and 
substrate temperature decreases with time as can be understood from fig 2(b) 
and (c).

If the heating and cooling is done in vacuum the cooling of the film, i.e.
loss of heat can take place by IR radiation losses or by conduction through 
the substrate side. After the heater is switched off, since the process is in 
vacuum, the substrate temperature remains constant for an appreciably long time 
before it starts falling. We assume that the fall in substrate temperature 
takes place after 200sec from instant that the heater is switched off. Due to 
the temperature gradient present along the thickness of the film, the surface 
tends to attain the same temperature as the substrate. Figure 3 shows the    
temperature along the thickness of the  film at various time, between the
instant when the heater was switched off  to 200 sec after it was switched off. 
It can be seen from figure 3 (a), the spatial distribution of temperature 
remains the same even after 200 sec due to the poor thermal conductivity of 
the film. However, as can be seen from fig 3(b) and (c), with improving thermal 
conductivity, the film eventually attains equilibrium with time.

The variation of the substrates temperature with time is taken as  
\begin{eqnarray}
T_{sub} = Se^{-ut} + R\nonumber
\end{eqnarray}
where S is the maximum temperature that the substrate had reached in the
heating cycle, before the onset of cooling. For computation of the temperature
distribution, to maintain assumption that the rate of cooling is very different 
from the heating rate, 'u' is taken  as ${\rm 0.00012sec^{-1}}$. Figure 4 shows 
the temperature distribution along the length of the film at regular intervals 
after the onset of cooling. For the film of low thermal conductivity (fig 4 a), 
the distribution profile is very similar to that of heating cycle as shown in 
fig 3 (a). However, as exhibited by figure 4 (b) and (c), the profile is 
different for films with better thermal conductivity, where in some cases
(4c, ii-v), the surface is seen to at a higher temperature then the substrate. 
This immediately suggests there would be some difference in film properties, 
such as resistance, during heating and cooling cycle.

\section{Film Resistance}

The film can be thought of an numerous infinitesimal identical thin layers,
one on top of the other. All the layers acting as resistive elements with the
net resistance of the film being the effect of these resistance appearing in
parallel combination. Since the layers are identical, at room temperature all 
of them have equal value. However, due to the metallic/ semiconducting nature 
of the film, the resistance of these layers vary with temperature. The variation 
of resistance with temperature is given as
\begin{eqnarray}
R = R_o(1+ \alpha T)\nonumber
\end{eqnarray}
where ${\rm \alpha}$ and ${\rm R_o}$ are the temperature coefficient of 
temperature (TCR) and the resistance of the identical layer. For the case ${\rm
T=0^oC}$, the films resistance would be given as  
\begin{eqnarray}
{1 \over R} = \sum_{i=1}^{i=n} {1 \over R_o} = {n \over R_o} 
\end{eqnarray}
The TCR is positive for metal while it is negative for semiconductors. Since, 
spatial distribution of temperature was calculated for various substrate 
temperatures at various instant, the films resistance can be trivially 
calculated as a function of substrate temperature and time. 

Figures 5-7 were plotted with data generated assuming the 1000\AA\, thick film
to be made up of  10 resistive layers in parallel combination, with each layer 
to have a room temperature resistance of ${\rm 170 K\Omega}$ and ${\alpha= -0.80 
\time 10^{-3o}C^{-1}}$. These numerical values are taken from our previous
study on ${\rm Sb_2Te_3}$ films \cite{7}. Figure 5 shows the variation of 
resistance with substrate temperature. As can be seen films with 
moderate thermal conductivity and those with good thermal conductivity
enclose very small area. However, films with intermediate diffusivity enclose 
large area due to aggravated difference between the heating and cooling cycle.

Figure 7 is of interest. The TCR or the variation of resistance with
temperature has been calculated for various diffusivity. It is evident that
the TCR of good thermally conducting films match the TCR of it's constituent 
infinitesimal thin layer of which the film is made of. For a mathematical
analysis consider the film to be made up of infinite strips of layer, such that 
each neighbouring layer has a slightly different temperature and inturn a
slightly different resistance. The summation sign of equation (6) may then
be replaced by an integration sign, hence the net resistance of the film
would be given as
\begin{eqnarray}
{1 \over R} = {1 \over R_o}\int_{i=0}^{n=d/a} {di \over (1 + \alpha
T)}\nonumber
\end{eqnarray}
At an given instant the temperature is given by equation (5), hence the
above equation can be re-written as
\begin{eqnarray}
{R_o \over R} = \int_{0}^{n} {di \over (1 + \alpha T_{sub})
-\alpha(T_{sub}-T_{sur})e^{-{\pi^2 D t \over 4 d^2}} sin \left ( {\pi \over 2
n}i \right )}
\end{eqnarray}
For solving the above equation, we substitute
\begin{eqnarray}
A &=& 1 + \alpha T_{sub}\nonumber \\
B &=& \alpha(T_{sub}-T_{sur})e^{-{\pi^2 D t \over 4 d^2}}\nonumber \\
x &=& {\pi \over 2 n}i\nonumber
\end{eqnarray}
Thus, equation (7) can be written as
\begin{eqnarray}
{\pi R_o \over 2 nR} = \int_{0}^{\pi /2} {di \over A- Bsin \left ( {\pi \over 2
n}i \right )}
\end{eqnarray}
\begin{eqnarray}
{\pi R_o \over 2 n R} = {1 \over \sqrt{A^2-B^2}} tan^{-1} \left ( \sqrt{{A+B
\over A-B}} \right )
\end{eqnarray}
As the film's diffusivity increases, the term B becomes smaller and
smaller, i.e. tends to zero. The above equation then reduces to
\begin{eqnarray}
{\pi R_o \over 2 n R} = {1 \over A} tan^{-1}(1)= {\pi \over 2A}\nonumber 
\end{eqnarray}
or
\begin{eqnarray}
{nR \over R_o} = 1+\alpha T\nonumber 
\end{eqnarray}
on re-arranging
\begin{eqnarray}
R = {R_o \over n}(1+\alpha T)\nonumber 
\end{eqnarray}
using the equation showing rise in temperature with time, we have
\begin{eqnarray}
{dR \over dt} = {R_o \over n}(\alpha PQe^{-Qt})\nonumber 
\end{eqnarray}
and
\begin{eqnarray}
{dT \over dt} = PQe^{-Qt}\nonumber 
\end{eqnarray}
Thus the film's TCR works out as
\begin{eqnarray}
TCR_{film} = {n \over R_o}{dR \over dT} = \alpha\nonumber 
\end{eqnarray}
The mathematics show that not only does a good thermally conducting film's TCR 
match that of the infinitesimal thin layer of which the film is made of, it
is also independent of the rate of heating/ cooling. It can be inferred that 
the TCR of the film's with lower thermal conductivity would show dependence on 
the rate of heating and cooling. This can be seen from figure 8. Figure 8 shows 
the effect the rate of heating would have on the slope, and
inturn the TCR. The data was calculated in the same manner as
discussed in the previous sections. While Figure 8A exhibits the variation
of resistance with temperature for a poor thermally conducting film
(D=500\AA/sec), Figure 8B is for a good conducting film (D=5000\AA/sec).
Three curves are present in both figures, each for different heating rates,
namely (i) ${\rm 3.6 \times 10^{-3o}C/sec}$, (ii) ${\rm 72 ^oC/sec}$ and 
(iii) ${\rm 216 ^oC/sec}$. All three curves coincide for the conducting film. 
However, in the figure 8A, where a low thermal conducting film the curves do 
not coincide and their slopes are different. Thus, the TCR values would depend 
on the rate of heating and cooling. An interesting feature is that the
resistances at various temperatures of a poor conducting film  measured at
very low heating rates match those of a good conducting film being heated
rapidly.

\section{Conclusions}
The electrical studies of thin films are usually done by heating the sample
and measuring resistance/ resistivity with temperature. Though, the
measurements are to be done after the film has attained a steady
temperature, usually the measurement is done as the film is being heated or
cooled. As discussed in the article, if the film has a finite thermal
conductivity (i.e. it is not metallic), one essentially is making
measurement in non-equilibrium conditions. Thus, parameters like TCR etc.
computed is not only material dependent but depends on conditions of the
experiment, e.g. the rate of heating or cooling. It is essentially due to
this non-equilibrium measurement that leads to a loop like formation due to
the heating-cooling cycle.

\begin{acknowledgments}
The authors would like to acknowledge the contributions of Prof. S. R.
Choudhury, Pankaj Tyagi and Naveen Gaur.
\end{acknowledgments}

\pagebreak

\pagebreak

\begin{figure}
\begin{center}
\epsfig{file=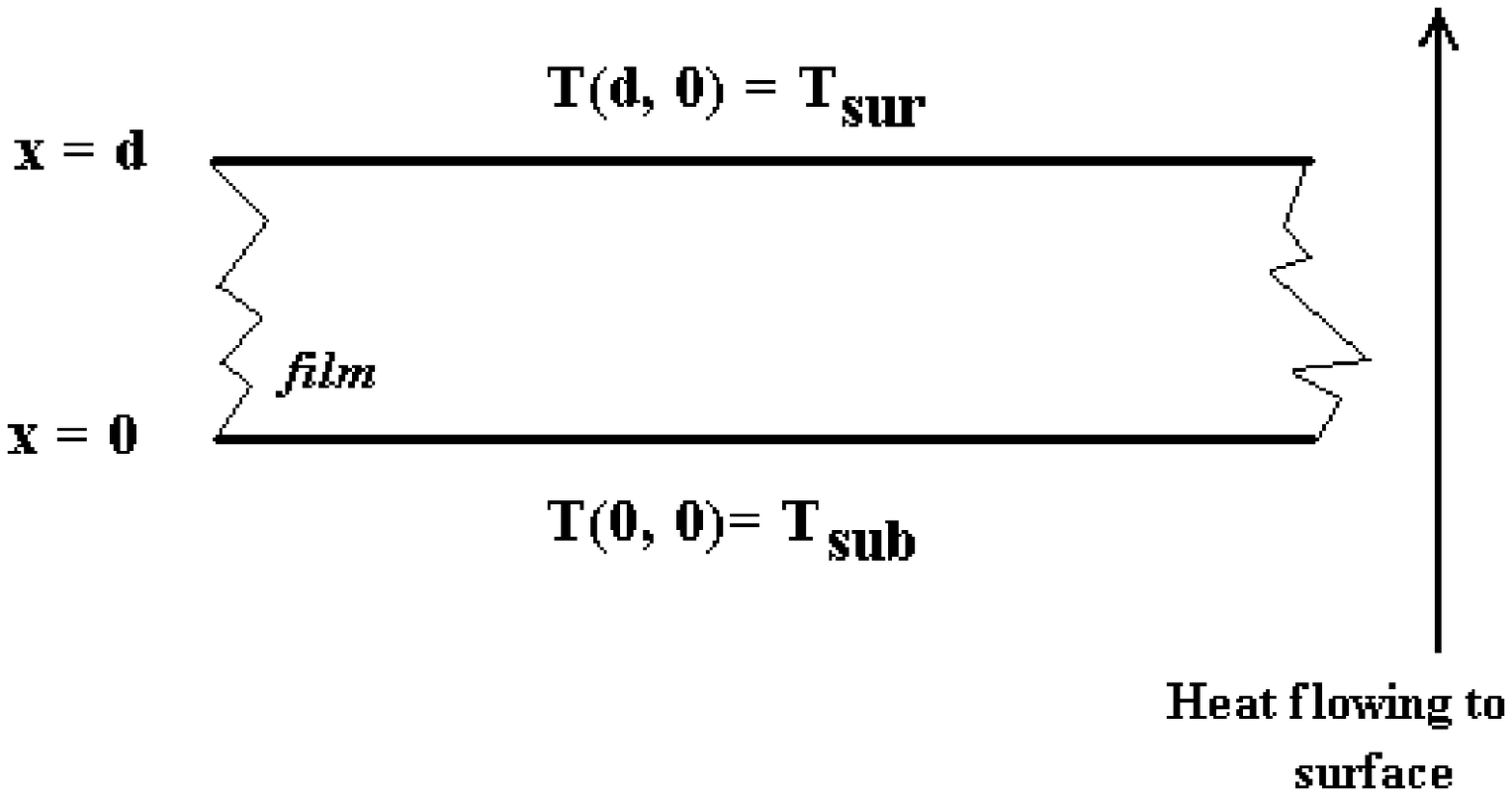,width=5 in}
\caption{ Direction of heat flow and initial condition of temperature on both 
surfaces of the film.
} 
\label{fig:1}
\end{center}
\end{figure}

\begin{figure}
\begin{center}
\epsfig{file=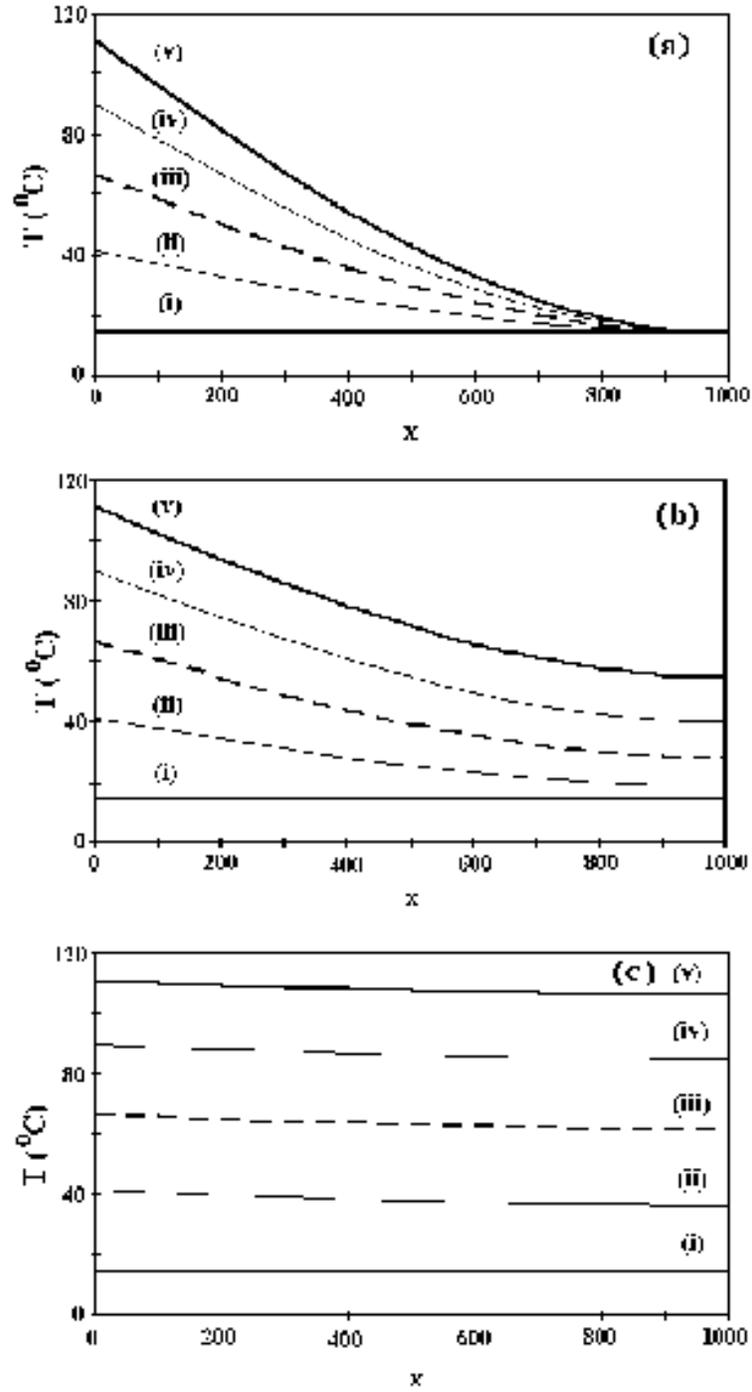,height=7.5in, width=4 in}
\caption{Variation of temperature along the thickness of a 1000\AA\, thick
film of different diffusivity (a) ${\rm 5 \times 10^{-3}\AA/sec}$, 
(b) ${\rm 5 \times 10^{2}\AA/sec}$ and (c) ${\rm 5 \times 10^{3}\AA/sec}$ after 
(i) 0 sec, (ii) 200 sec, (iii) 400 sec, (iv) 600 sec
and (v) 800 seconds of substrate heating.} 
\label{fig:2}
\end{center}
\end{figure}

\begin{figure}
\begin{center}
\epsfig{file=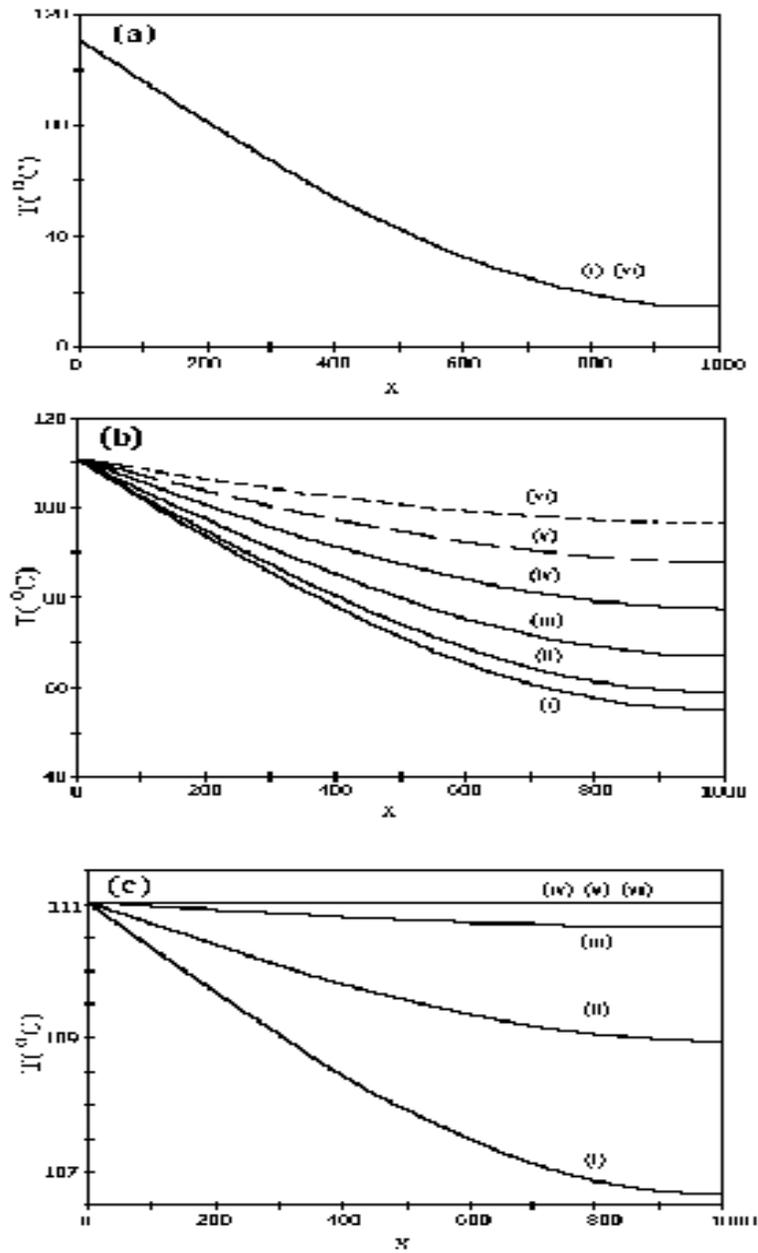,height=6.75in, width=4 in}
\caption{Variation of temperature along the thickness of a 1000\AA\, thick
film of different diffusivity (a) ${\rm 5 \times 10^{-3}\AA/sec}$, 
(b) ${\rm 5 \times 10^{2}\AA/sec}$ and (c) ${\rm 5 \times 10^{3}\AA/sec}$ after 
(i) 0 sec, (ii) 40 sec, (iii) 80 sec, (iv) 120 sec, (v) 160 sec
and (vi) 200 seconds after the source of heating was switched off.}
\label{fig:3}
\end{center}
\end{figure}

\begin{figure}
\begin{center}
\epsfig{file=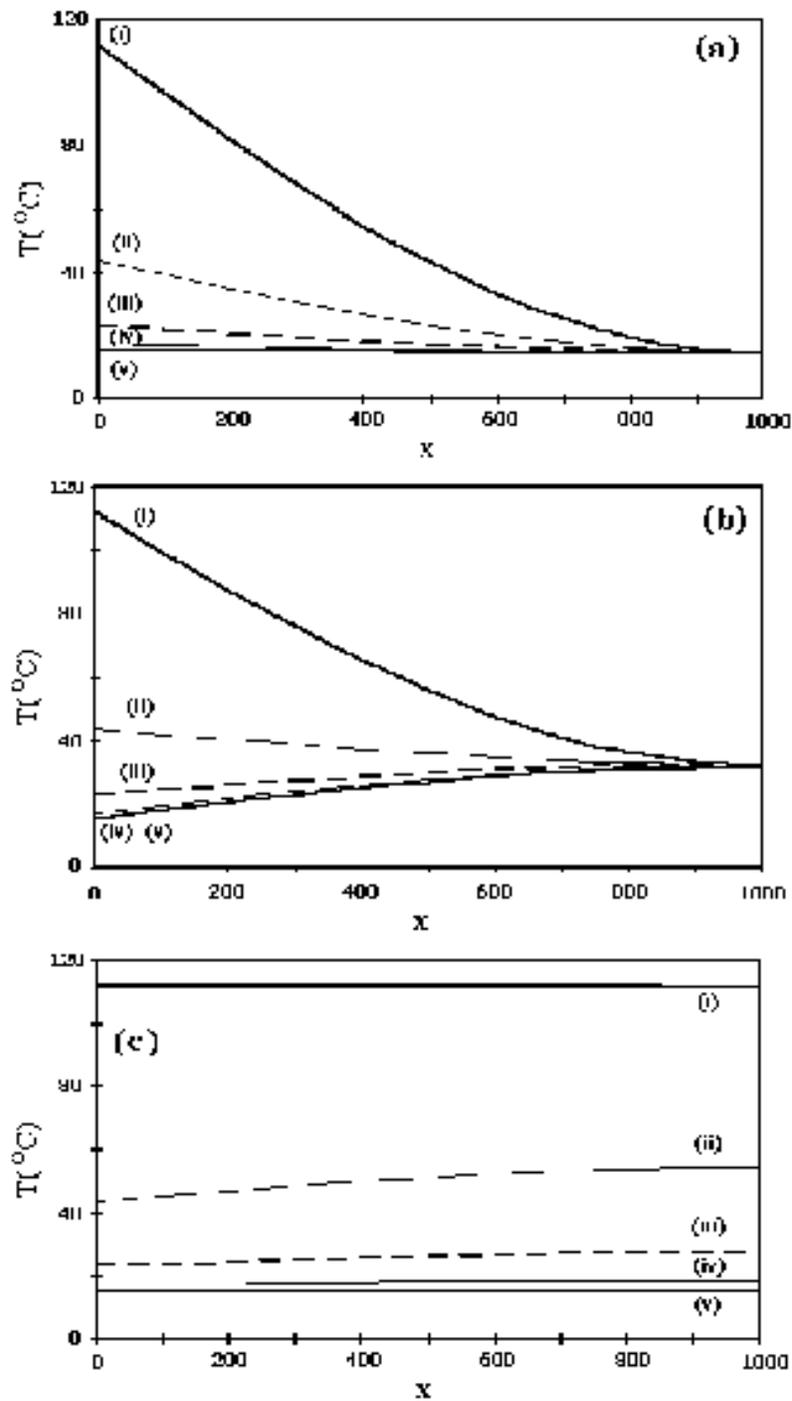, height=7.5in, width=4.2in}
\caption{ Variation of temperature along the thickness of a 1000\AA\, thick
film of different diffusivity (a) ${\rm 5 \times 10^{-3}\AA/sec}$, 
(b) ${\rm 5 \times 10^{2}\AA/sec}$ and (c) ${\rm 5 \times 10^{3}\AA/sec}$ after 
(i) 8000 sec, (ii) 16000 sec, (iii) 24000 sec, (iv) 32000 sec and (v) 40000 
seconds after 
the setting in of the film's cooling.}
\label{fig:5}
\end{center}
\end{figure}

\begin{figure}
\begin{center}
\epsfig{file=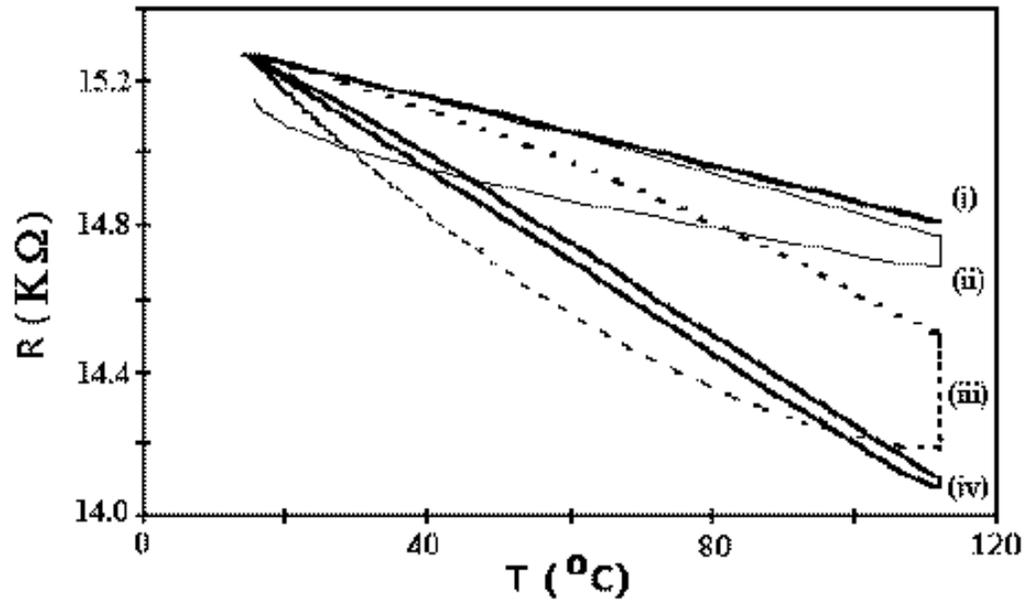,width=5.5 in}
\caption{  
Hysteresis loops formed in film resistance with the heating -cooling cycle.
The calculations were done for film thickness of 1000\AA\, and diffusivity (i)
${\rm 5 \times 10^{-3} \AA^2/sec}$, (ii) ${\rm 50 \AA^2/sec}$, (iii) ${\rm 5
\times 10^2 \AA^2/sec}$ and (iv) ${\rm 5 \times 10^3 \AA^2/sec}$.
}
\label{fig:5}
\end{center}
\end{figure}

\begin{figure}
\begin{center}
\epsfig{file=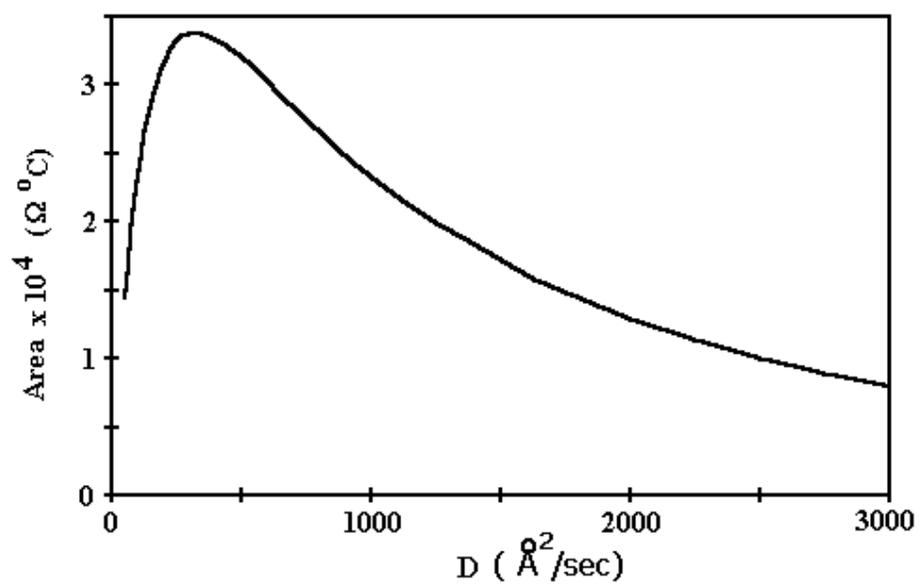,width=5 in}
\caption{ The variation in the area enclosed by loops formed during the
resistance variation with temperature during heating-cooling cycles. The
variation is due to the difference in the films diffusitivity. }
\label{fig:6}
\end{center}
\end{figure}

\begin{figure}
\begin{center}
\epsfig{file=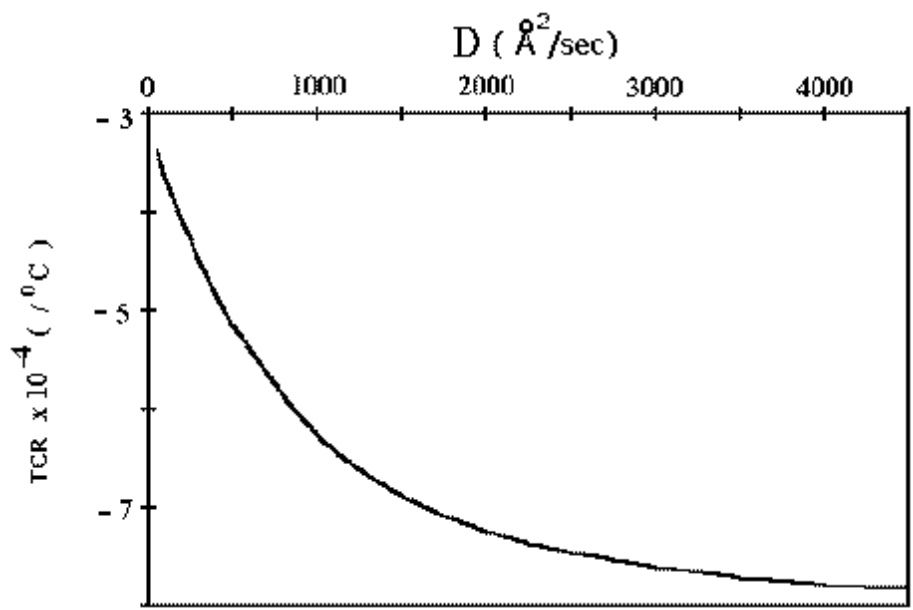,width=5 in}
\caption{ Computed TCR for films of different diffustivity, where the films
are assumed to be of same thickness and made up of numerous layers, with all
the layers having the same TCR.}
\label{fig:7}
\end{center}
\end{figure}

\begin{figure}
\begin{center}
\epsfig{file=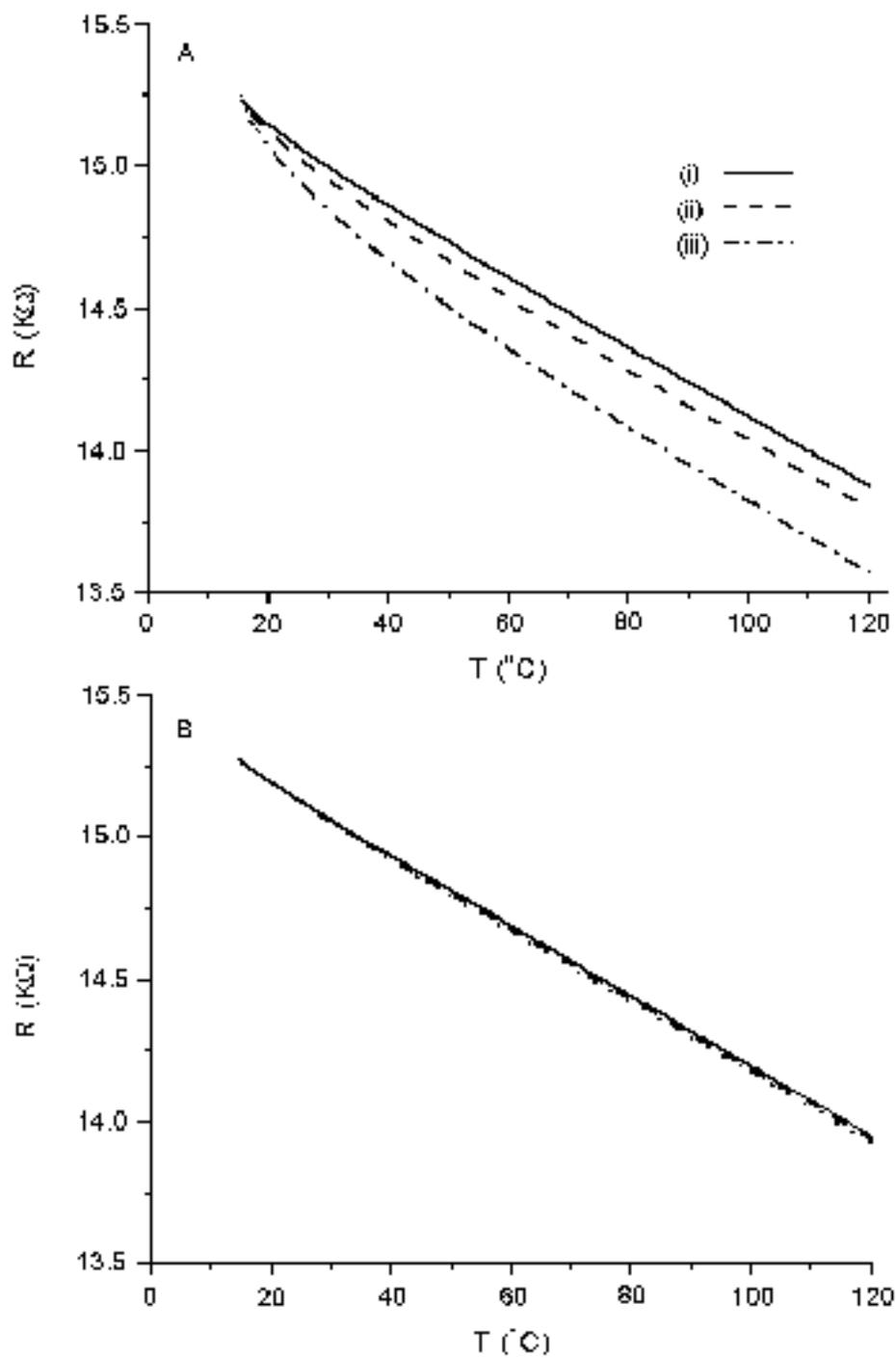,width=5 in}
\caption{ Figure exhibits the change of resistance of with temperature for
(A) a poor thermal conducting film and (D= ${\rm 5 \times 10^{2}\AA/sec}$) 
(B) a good thermally conducting film (D= ${\rm 5 \times 10^{3}\AA/sec}$).
The heating rates were maintained different (i) ${\rm 3.6 \times
10^{-3o}C/sec}$, (ii) ${\rm 72 ^oC/sec}$ and (iii) ${\rm 216 ^oC/sec}$.}
\label{fig:8}
\end{center}
\end{figure}
\end{document}